\documentstyle[preprint,aps]{revtex}
\begin{document}
\author{Yu.A. Firsov $^{1}$, V.V. Kabanov$^{2}$, E.K. Kudinov $^{1}$, and 
A.S. Alexandrov$^{3}$}
\address{
$^{1}$ A.F. Ioffe Physical-Technical Institute, Russian\\
Academy of Sciences, 194021 St.Petersbourg, Russia\\
$^{2}$ Bogoljubov Laboratory of Theoretical Physics,\\
JINR, Dubna, Russia; and Josef Stefan Institute SI-1001,\\
Ljubljana, Slovenia. $^{3}$ Department of Physics,\\
Loughborough University, Loughborough LE11 3TU, U.K. }
\title{Comment on `Dynamical properties of small polarons'}
\maketitle

\begin{abstract}
We show that the conclusion on the breakdown of the standard small polaron
theory made recently by E.V. deMello and J. Ranninger (Phys. Rev. B${\bf 55}$%
, 14872 (1997)) is a result of an incorrect interpretation of the electronic
and vibronic energy levels of the two-site Holstein model. The small polaron
theory, when properly applied, agrees well with the numerical results of
these authors. Also we show that their attempt to connect the properties of
the calculated correlation functions with the features of the intersite
electron hopping is unsuccessful.
\end{abstract}

\vskip0.5cm \noindent {\bf 1.}In a recent publication \cite{ran} de Mello
and Ranninger have analyzed numerically the familiar two-site Holstein model 
\cite{hol,firkud} of a single electron coupled to an intra-site vibration
mode. The model is the electronic doublet, $\phi _{1}$, $\phi _{2}$,
describing an electron legalized on sites $1$ and $2$, respectively, plus
the interaction with a vibration mode. The overlap of $\phi _{1}$ and $\phi
_{2}$ leads to a splitting $2t$ of the doublet in the absence of the
interaction, where 
\begin{equation}
t=\int \phi _{1}^{*}H_{{\rm el}}\phi _{2}\,dV,
\end{equation}
$H_{{\rm el}}$ is the electron Hamiltonian. The Hamiltonian of the model is
given by 
\begin{equation}
H={\frac{p^{2}}{{2M}}}+{\frac{M\omega _{0}^{2}x^{2}}{{2}}}-gx(a_{1}^{\dagger
}a_{1}-a_{2}^{\dagger }a_{2})-t(a_{1}^{\dagger }a_{2}+a_{2}^{\dagger }a_{1}).
\label{ham}
\end{equation}
Here $p,\,x$ are the momentum and co-ordinate of the vibration mode, $%
M,\,\omega _{0}$ its mass and frequency, respectively, and $a_{i}^{\dagger
},\,a_{i}$ $(i=1,2)$ are the electron operators. The Eq.(\ref{ham}) is an
invariant with respect to the inversion $x\to -x,\,(1,2)\to (2,1)$ and the
parity is conserved. This model is an extreme simplification of the small
polaron model. $2t$ is the analog of the electron bandwidth in a crystal.

\noindent {\bf 2.}One of the basic results of the small polaron theory \cite
{hol,lan,app,fir,fir1} is the narrowing of the electron band due to the
electron-phonon (vibron) coupling. In the strong-coupling regime, when $%
g^2\to\infty $, the narrowing is described by a simple exponential
dependence on $g$ as 
\begin{equation}
t^{*}=t \exp(-{\rm const}\cdot g^2).
\end{equation}
The exponential dependence, Eq.(2) is readily derived by the use of the
familiar double-well potential model \cite{hol}. Polaron tunnels coherently
within the narrow band at low temperatures (while the thermally activated
hopping dominates at high temperatures \cite{app,fir}). The numerical
solution for several vibrating molecules coupled with one or two electrons 
\cite{alekab,feha,kora,roma,irra} revealed an excellent agreement of the
numerical bandwidth with the analytical Holstein and Lang-Firsov (LF)
results at large $g$.

\noindent {\bf 3.}However, de Mello and Ranninger \cite{ran} arrived with an
opposite conclusion. On the basis of numerical analysis of the same problem
authors \cite{ran} claim in Sec.III that:

\begin{quote}
the LF approach\footnote{%
In Ref. \cite{ran} the "LF approach" is identified with the lowest ($\sim t$%
) order in perturbation theory. We do not agree with this identification,
since to go beyond the lowest approximation was a central point for the
authors of Ref. \cite{lan,fir,fir1}.}, which is generally believed to become
exact in the limit of antiadiabaticity and an electron-phonon coupling going
to infinity, actually diverges most from the exact results precisely in this
limit... (p. 14885)
\end{quote}

These authors did not provide any physical explanation for their drastic
disagreement with all earlier results starting from the pioneering work by
Holstein and including the kinetic theory of strongly-coupled
electron-phonon systems, in particular with the theory of high-frequency
conductivity \cite{fir,kun}.

\noindent {\bf 4.}In Ref. \cite{kudfir} we have recently developed the
analytical approach to the two-site model by the use of the expansion
technique, which provides the electronic and vibronic terms as well as the
wave functions in any order in powers of $t$. In the second order in $t$ the
doublet energy of the ground state, $E_{\pm }$ is given by 
\begin{equation}
E_{\pm }=\pm t^{*}-\frac{t^{2}}{4E_{p}},\quad t^{*}\equiv t\exp
(-2E_{p}/\hbar \omega _{0}).  \label{doub}
\end{equation}
Here $E_{p}=g^{2}/2M\omega _{0}^{2}$ is the polaron shift (the following
designation is used in Ref. \cite{ran}: $E_{p}=\alpha ^{2}\hbar \omega _{0}$%
). The first term describes the splitting of the doublet (components of the
doublet have opposite parity) corresponding to the bandwidth in a crystal,
as discussed above, while the second term is a correction to the polaron
shift of the whole band due to the virtual transitions to the
nearest-neighbour site. The exponential reduction factor was found in all
orders of $t$ of the perturbation expansion \cite{kudfir} in agreement with
the standard result, Eq.(2). On the other hand the corrections to the atomic
level are relatively small as $1/g^{2}$ rather than exponential.

\noindent {\bf 5.} In Ref. \cite{kudfir}, Sec.6 it was demonstrated, that
the aforesaid statements of authors of Ref. \cite{ran} do not correspond to
reality and is only due to the fact that they failed to notice the
above-mentioned difference between the splitting of the doublet's components
and their shift as a whole. This is the result of the methodological defect
of the approach which was employed in Ref. \cite{ran}. Instead of a direct
solution of the quantum-mechanical problem (to determine the energy spectrum
and the wave functions \footnote{%
This approach would eliminate the possibility to make such an error to a
considerable extent.}), they calculated a value (which has no direct
physical meaning) 
\begin{equation}
E_{{\rm kin}}^{i}\sim t\langle a_{1}^{\dagger }a_{2}\rangle _{i},
\end{equation}
$\langle \dots \rangle _{i},\,(i=+,\,-)$ is a quantum-mechanical average on
the one of the doublet's component. They accepted implicitly an assertion
that:

\begin{quote}
$E_{{\rm kin}}^{i}$ is the analog of the electron bandwidth \footnote{%
No explicit wording of this assertion is given in Ref. \cite{ran}. However,
this wrong interpretation of $E_{{\rm kin}}^{i}$ is forced, because the
authors of Ref. \cite{ran} have compared it just with a small polaron
bandwidth \cite{hol,lan}. In other case such comparsion would be irrelevant,
since authors of Ref. \cite{hol,lan} never calculated $E_{{\rm kin}}$.}.
\end{quote}

As it was shown in Ref. \cite{kudfir}, this assertion is incorrect. Let us
denote as $\Delta E^{i}$ a correction term to the energy level $i$ which is
generated by the last term $\sim t$ of the Hamiltonian (2). In fact:

\begin{enumerate}
\item  $E_{{\rm kin}}^{i}$ is proportional to $\partial \Delta
E^{i}/\partial t$ (rather then to $\Delta E^{i}$). The average (5) decreases
when $|g|$ increases as a power of $g^{2}$ ($\sim g^{-2}$ for $g^{2}\to
\infty $, see Ref. \cite{alemot}), but not exponentially!

\item  Analog of the electron bandwidth is the difference $\delta E=|\Delta
E^{+}-\Delta E^{-}|$, but not $\Delta E^{i}$ alone. $\delta E$ contains the
exponential factor (3).
\end{enumerate}

\noindent In other words, in Ref. \cite{ran} authors had compared the values
of the essential different nature. This is the source of the above-mentioned
drastic disagreement.

\noindent {\bf 6.} In Ref. \cite{ran}, Sec.V authors have calculated the
electron dipole momentum\footnote{%
The authors of Ref. \cite{ran} have connected this correlator with the
charge fluctuations which is not quite correct.} and the vibronic
co-ordinate correlators (deformation dynamics correlator in Ref. \cite{ran}%
): 
\begin{equation}
\chi _{{\rm nn}}(\tau )=\langle (n_{1}-n_{2})_{\tau }(n_{1}-n_{2})\rangle
_{0},\quad \chi _{{\rm xx}}(\tau )=\langle x_{\tau }x\rangle _{0},
\end{equation}
where $\langle \dots \rangle _{0}$ is the average over the ground state $%
\Psi _{0}$. Not any analytical examination was made. The authors of Ref. 
\cite{ran} pointed out that the calculated curves which represent a
functional dependence $\chi (\tau )$ (Figs. 11, 12 of Ref. \cite{ran}) may
be presented as a superposition of a slow and fast oscillations with the
frequencies of the fast oscillations $\widetilde{t}$ and $\widetilde{\omega }
$ accordingly for $\chi _{{\rm nn}}(\tau )$ and $\chi _{{\rm xx}}(\tau )$ 
\footnote{%
We do not see any reason to identify $\widetilde t$ with "renomalized
intrisinc hopping integral $t$".}. They claim that when these frequencies
(which are certainly $>\omega _{0}$, see Table 1 in Ref. \cite{ran}) draw
together, the qualitative changing of the electron transport mechanism takes
place. No physical argumentation (even of a qualitative nature) to support
this assertion of the authors have been given.

\noindent {\bf 7.} We note that $\chi (\tau )$ may be represented as 
\begin{equation}
\chi (\tau )=\sum_{m\neq 0}a_{m}^{2}e^{-i\omega _{m0}\tau },\quad \hbar
\omega _{m0}=E_{m}-E_{0},\quad a_{m}=\langle \Psi _{m}A\Psi _{0}\rangle ,
\label{chi}
\end{equation}
$\Psi _{m},\,E_{m}$ are eigen-functions and eigenvalues, accordingly, of the
Hamiltonian Eq.(\ref{ham}), $A=n_{1}-n_{2}$ or $x$ . The summation In (\ref
{chi}) is performed over the states $\Psi _{m}$ with parity opposite to the
parity of the ground state $\Psi _{0}$ (selection rules for the operator $A$%
). We note that the frequency spectra of $\chi _{{\rm nn}}$ and $\chi _{{\rm %
xx}}$ are identical.

There is a connection between correlators (6), (7), and corresponding
generalized susceptibilities $\kappa _{a}(\omega )$ (see Ref. \cite{land}).
Here $\omega $ is the frequency of an external disturbance. For example, the
complex polarizability $\kappa (\omega )$ of the considered model may be
expressed by the Fourier-transform of the correlator (6) 
\begin{equation}
\kappa (\omega )=\frac{ie^{2}l^{2}}{\hbar }\int_{0}^{\infty }e^{i(\omega
+i\delta )\tau }\left( \chi _{{\rm nn}}(\tau )-\chi _{{\rm nn}}(-\tau
)\right) \,d\tau ,\quad \delta >0,\,\delta \to 0.
\end{equation}
$l$ is a constant with dimensionality of the length. An imaginary part of $%
\kappa (\omega )$ is 
\begin{equation}
\kappa ^{^{\prime \prime }}(\omega )\!=\!\frac{ie^{2}l^{2}}{\hbar }%
\sum_{m}a_{m}^{2}\left( \delta (\omega -\omega _{m0})-\delta (\omega +\omega
_{m0})\right) .  \label{abs}
\end{equation}
The value $\omega \kappa ^{^{\prime \prime }}(\omega )$ determines an
absorption coefficient of electromagnetic radiation, and the value $\omega
_{m}a_{m}^{2}$ determines an absorption intensity for the transition $0\to m$%
. Due to the presence of the $\delta -$functions in Eq.(\ref{abs}), the
absorption process in the given frequency range can not be linked causally
with another one in the other frequency range \footnote{%
For example, a DC conductivity of the semiconductor does not depend on a
higher empty band contribution (and vice versa, the intensity of the
interband absorbtion does not depend on DC conductivity mechanism).}.

It is natural to identify the frequencies in Ref. \cite{ran}, Figs. 11, 12
in the following way: the slow oscillation corresponds to the width $\delta
E=2t^{*}$ of the lowest doublet, Eq. (\ref{doub}); the fast oscillations
correspond the frequencies $>\delta E/\hbar $ in $\chi _{{\rm nn}}(\tau )$
and $\chi _{{\rm xx}}(\tau )$ for that the weights $a_{m}^{2}$ (see Eq.(\ref
{chi})) are maximal. The slow and fast oscillations are located in different
frequency regions, therefore changes in the high-frequency region cannot
modify the low-frequency electron transport mechanism cardinally.

For these reasons, the aforesaid assertion (see above, Sec. 6) is unfounded.
And again, the source of this error is of methodological character. This
situation would be excluded, if instead of $\chi (\tau)$, which have no
direct physical meaning, they have considered the complex polarizability $%
\kappa(\omega)$.

In Ref. \cite{ran} the authors touch upon the subject of a boundary where a
localized regime changes to an itinerant one. In our opinion the boundary is
determined by the parameter $\eta_1=t/2E_p$ \footnote{$\eta_1$ is the
parameter which was introduced by T.Holstein; the small polaron appears when 
$\eta_1<1$. Also this parameter determines a correction ($\sim t$) to the
overlap integral between the site-localized functions.}. For $\eta_1<1$ the
lowest adiabatic potential curve has two minima, which are separated by the
energetic barrier (localized regime), for $\eta_1>1$ the barrier vanishes
(itinerant regime), Ref.\cite{kudfir}. We state, that in the range of the
parameters, which considered in Ref. \cite{ran} the itinerant regime was not
yet realized.

\noindent {\bf 8.} We have checked and proved that (under right
interpretation, naturally) the numerical calculations presented in Sec. III
and V of Ref. \cite{ran} agree satisfactorily with the Holstein-LF approach
for $t/\hbar \omega _{0}<1$\footnote{%
The deviations which arise when $t/\hbar\omega_0\ge1$ may be explained
qualitaively in the framework of the adiabatic approach.}.

Finally, we note that the authors' assertion in Sec. V \cite{ran}{\ }

\begin{quote}
`We notice that the charge dynamics qualitatively tracks globally the
behavior expected on the basis of the LF approximation in the anti-adiabatic
limit... `(p.14882) 
\end{quote}
obviously clashes with their statement in Sec.. III (see above, the
quotation in our sec.3 ). No comments on this discrepancy are given in Ref. 
\cite{ran}. \vskip0.5cm \noindent In conclusion we state that although the
numerical calculations in Ref. \cite{ran} were performed fairly enough, yet
their interpretation is untenable.

One of us (V.V.K.) acknowledges support of the work by RFBR Grant 97-2-16705
and the Ministry of Science and Technology of Slovenia.


\begin{references}
\bibitem{ran}  E.V.L. de Mello and J. Ranninger, Phys. Rev. B ${\bf 55}$,
14872 (1997).

\bibitem{hol}  T. Holstein, Ann.Phys. ${\bf 8}$, 325-42; ibid p. 343 (1959).

\bibitem{firkud}  E.K.Kudinov and Yu.A.Firsov, Fiz. Tverd. Tela, {\bf 7},
546, (1965); (In English: Soviet Physics-Solid State {\bf 7}, 435 (1965)).

\bibitem{lan}  I.G. Lang and Yu.A. Firsov, Zh.Eksp.Teor.Fiz. ${\bf 43}$,
1843 (1962) ( Sov.Phys.JETP ${\bf 16}$, 1301 (1963)).

\bibitem{app}  J.Appel, in Solid State Physics, eds. F. Seitz, D. Turnbull
and H. Ehrenreich, Academic Press ${\bf 21}$ (1968).

\bibitem{fir}  
Yu.A. Firsov (ed), Polarons, Nauka (Moscow) (1975).

\bibitem{fir1}  Yu.A. Firsov, Semiconductors, {\bf 29}, 515 (1995).

\bibitem{alekab}  A.S. Alexandrov, V.V. Kabanov and D.K. Ray, Phys.Rev. B$%
{\bf 49}$, 9915 (1994).

\bibitem{feha}  H. Fehske, J. Loos, and G. Wellein, Z. Phys. B I. ${\bf 104}$%
, 619 (1997).

\bibitem{kora}  P.E. Kornilovitch and E.R. Pike, Phys. Rev. B ${\bf 55}$,
R8635 (1997).

\bibitem{roma}  A.H. Romero, D.W. Brown and K. Lindenberg, Cond-mat/9710321
1997).

\bibitem{irra}  E. Jeckelmann and S. White, Phys. Rev. B {\bf 57}, 6376
(1998).

\bibitem{kun}  Huang Kun, A.Rhys. Proc.Roy.Soc. {\bf A208},352 (1951).

\bibitem{kudfir}  
Yu.A. Firsov and E.K. Kudinov, Fiz. Tverd. Tela, {\bf 39}, 2159 (1997);(in
English: Phys. Solid. State (AIP) {\bf 39}, (12), 1930 (1997)).

\bibitem{alemot}  A.S. Alexandrov and N.F. Mott, Rep.Prog.Phys. ${\bf 57}$,
1197 (1994).

\bibitem{land}  L.D. Landau and E.M. Lifshitz. Statistical Physics,
Gostehizdat (Moscow), in Russian (1951).
\end{references}
\end{document}